\documentclass[a4paper,fleqn]{cas-dc}
\usepackage[utf8]{inputenc}
\usepackage[numbers]{natbib}
\usepackage[T1]{fontenc}
\usepackage{lineno}
\usepackage{hyperref}
\usepackage{hyphenat}
\modulolinenumbers[5]

\usepackage[dvipsnames]{xcolor}
\usepackage{xspace}
\newcommand\chaosorca{{\sc ChaosOrca}\xspace}
\newcommand\chaosorcabf{{\sc \textbf{ChaosOrca}}\xspace}

\usepackage{csvsimple}
\usepackage{longtable}
\usepackage{lscape}
\usepackage{array}
\csvstyle{csvResultStyle}{
  head to column names,
  tabular=|p{.18\linewidth}|p{.25\linewidth}|p{.26\linewidth}|p{.36\linewidth}|,
  table head=\hline \bfseries System call perturbation & \bfseries HTTP requests & \bfseries System calls & \bfseries Other observations \\ \hline,
  late after last line=\\\hline
}
\usepackage{subfig}

\usepackage{array, etoolbox}
\newcounter{rowcount}
\usepackage{amsthm}
\theoremstyle{definition}
\newtheorem{requirement}{Requirement}
\newtheorem{question}{RQ}

\usepackage[framemethod=tikz]{mdframed}
\mdfdefinestyle{mpdframe}{
    frametitlebackgroundcolor   =black!15,
    frametitlerule              =true,
    roundcorner                 =1pt,
    middlelinewidth             =1pt,
    innermargin                 =0.1cm,
    outermargin                 =0.1cm,
    innerleftmargin             =0.1cm,
    innerrightmargin            =0.1cm,
    innertopmargin              =0.1cm,
    innerbottommargin           =0.1cm
}

\usepackage{algorithm}
\usepackage{algorithmic}

\AtBeginDocument{%
}

\newcommand{\reviseadd}[1]{#1}

\begin{document}

\hyphenation{cha-os-or-ca}
\hyphenation{con-tain-er}

\let\WriteBookmarks\relax
\def\floatpagepagefraction{1}
\def\textpagefraction{.001}
\newcommand\mytitle{Observability and Chaos Engineering on System Calls for Containerized Applications in Docker}
\shorttitle{Observability and Chaos Engineering on System Calls}
\shortauthors{Simonsson et~al.}

\title [mode = title]{\mytitle}

\author[kth-address]{Jesper Simonsson}\ead{jsimo@kth.se}
\author[kth-address]{Long Zhang}\ead{longz@kth.se}
\author[sintef-address]{Brice Morin}\ead{Brice.Morin@sintef.no}
\author[kth-address]{Benoit Baudry}\ead{baudry@kth.se}
\author[kth-address]{Martin Monperrus}\ead{martin.monperrus@csc.kth.se}

\address[kth-address]{KTH Royal Institute of Technology, SE-100 44 Stockholm, Sweden}
\address[sintef-address]{SINTEF, Pb. 124 Blindern, 0314 Oslo, Norway}

\begin{abstract}
In this paper, we present a novel fault injection system called \chaosorca for system calls in containerized applications. \chaosorca aims at evaluating a given application's self-protection capability with respect to system call errors. The unique feature of \chaosorca is that it conducts experiments under production-like workload without instrumenting the application. We exhaustively analyze all kinds of system calls and utilize different levels of monitoring techniques to reason about the behaviour under perturbation. We evaluate \chaosorca on three real-world applications: a file transfer client, a reverse proxy server and a micro-service oriented web application. Our results show that it is promising to detect weaknesses of resilience mechanisms related to system calls issues.
\end{abstract}

\newcommand\contriba{A novel architecture for achieving observability and system call perturbation for containerized applications in Docker, able to analyze the resilience of Docker applications in a black-box manner.}
\newcommand\contribb{A prototype implementation based on the state-of-art libraries used in industry, applicable to off-the-shelf Docker microservices. The prototype is called \chaosorca and is made publicly available at \url{https://github.com/KTH/royal-chaos/tree/master/chaosorca}.}
\newcommand\contribc{An evaluation of the approach on three case studies, representing typical polyglot microservices in Docker, namely TTorent (written in Java), Nginx (written in C) and Bookinfo (written in Java, JavaScript, Python and Ruby)}


\maketitle

\section{Introduction}

Docker containers are increasingly getting adopted as an efficient mean to package and deploy applications~\cite{dockerRelease2013Local}: over two million container images are available on DockerHub and popular images are downloaded millions of times. A typical system is composed of a number of Docker containers interacting with each others. Each container is an instance of an image, which describes the software stack that is needed by the application, starting from the OS, for example \texttt{debian} or \texttt{alpine}, all the required dependencies, and so on. The compiled application and configuration files are copied in the image. The images used in production are typically optimized for performances and security, meaning that anything that is not specifically required by the application is stripped-down: in the Docker image there is no \texttt{ssh} access to log into the container, no remote debugger, and so on.

Docker is a key-enabling technology for microservices, an architectural style promoting simple, loosely-coupled, individually updateable services organized around business products \cite{DockerDocumentationLocal,Jaramillo2016SECON}. Rather than imposing a generic framework that all microservices need to implement, a key principle of microservices is that different microservices share no code: each microservice is implemented using the technological stack that is best suited. This stack is then specified as a portable Docker image. As a result, different microservices significantly differ one from another: they can be implemented in different programming languages, use different libraries, and so on. As Docker containers rely on the kernel of the host OS, the only thing they all have in common is the kernel. Thus, they all rely on system calls to delegate certain actions to the guest operating system (OS) kernel, such as networking or writing to a file. 

Docker and microservices bring many benefits, in particular increasing the rate at which changes can be brought to production: for example, Netflix releases hundreds of changes to production every day. However, while the state of a monolithic system is rather binary, it either works according to the SLA or not, the state of a complex system built around microservices emerges from their interactions. In a monolithic system, a common framework or middleware would typically take care of providing reliable communications between components. This is no more the case with microservices, where each microservice needs to cope with the failures of other microservices and with the failures of the underlying stack. In this context, how to observe, analyze and improve the resilience of systems built over Docker and microservices?

This paper presents \chaosorca, a tool to conduct chaos engineering~\cite{Basiri2016} experiments to assess the resilience of {\em any} Docker-based microservices. Following the principles of chaos engineering~\cite{Principles_Of_Chaos_Engineering}, \chaosorca supports the following.

First, \chaosorca supports formalizing the steady state of the container, by automatically recording system metrics such as CPU and RAM consumption, network I/O and the system calls invoked by the container. In addition, \chaosorca supports additional metrics provided by developers.
Second, \chaosorca supports triggering events than can impact the steady state of the container, by injecting perturbations into the system calls invoked by the containers. For a given system call invoked by the container, a perturbation can delay the actual execution of the system call or can force returning an error code.
Third, \chaosorca enables developers to specify experiments on the production system. By default, \chaosorca perturbs the $N$ most invoked system calls for a given container, and monitors it before, during and after the experiment. It then provides a report to the developers summarizing the experiment, for them to verify of falsify the resilience hypothesis for that container.
Finally, \chaosorca is built on isolation, it allows experiments to target specific containers, while the other containers running aside are not be impacted by the injected perturbations.

A key feature of \chaosorca is that it is oblivious from the internals of the Docker containers involved in chaos experiment. In other words, no matter what software stack runs inside the container, \chaosorca will always be able to involve that container in a chaos experiment. Our empirical evaluation on TTorent (written in Java), Nginx (written in C) and Bookinfo (written in Java, JavaScript, Python and Ruby) shows that \chaosorca 1) can observe Docker-based microservices of very different nature, 2) can identify issues related to the resilience of those microservices against perturbation in system calls, 3) can improve the observability of microservices and the relevance of chaos experiments and 4) it can be applied in production with low overhead, though not all kinds of overhead are equal.

To sum up our contributions are:

\begin{itemize}
\item \contriba
\item \contribb
\item \contribc
\end{itemize}

The remainder of the paper is organized as follows. \autoref{sec:background} gives a brief introduction about relevant concepts. \autoref{sec:design} discusses about the design of \chaosorca like the requirements, components and implementation. \autoref{sec:evaluation} answers the 4 research questions by discussing about the evaluation experiments. \autoref{sec:relatedwork} compares the related work in fault injection and observability. \autoref{sec:conclusion} concludes the whole paper.

\section{Background}\label{sec:background}

\subsection{System Call}

A system call is the fundamental interface between an application and the kernel~\cite{syscallsmanLocal}. In the Linux operating system, there are more than 300 unique system calls and more than 100 different error codes that can be returned from system calls~\cite{LinuxSystemCalls}.
System calls are primitives exposed by the OS kernel to deal with crucial resources like hardware devices, network or files in a systematic and secure manner. When an application requests the use of such a resource, it has to interact with the OS kernel through a system call.

\begin{figure}
    \centering
    \includegraphics[width=0.9\columnwidth]{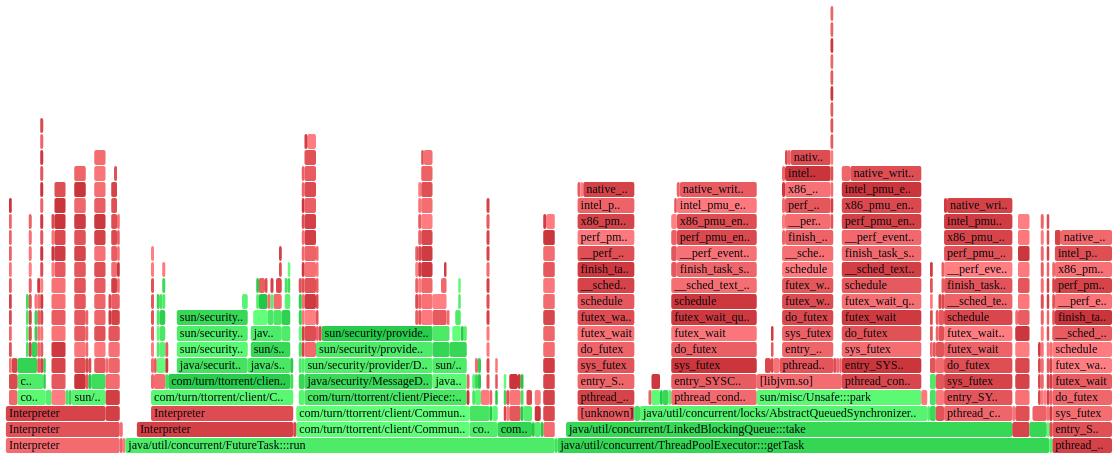}
    \caption{Flamegraph from TTorrent Execution. The green part shows application code, the red part shows OS code. \reviseadd{The width of the bars shows the frequency at which a function is present in the stack traces.}}
    \label{fig:flamegraph}
\end{figure}

It is very common for applications to invoke system calls, even if developers are not aware of such invocations. \autoref{fig:flamegraph} is a flame graph \cite{Gregg:2016} that illustrates user events and kernel events when downloading a file with TTorrent. In the figure, the green (light) blocks stand for user events, caused by Java code. The red (dark) ones are events related to system calls inside the kernel. The width of the bars stand for the frequency at which that function is present in the stack traces, or part of a stack trace ancestry.\cite{Gregg:2016} For TTorrent, those system calls mostly happen when it opens a file, uses network resources or writes downloaded blocks into a file. 

\subsection{Containerization}\label{sec:bg-containerization}
Containerization is a lightweight virtualization mechanism to support rapid shipment and deployment of applications. A containerized application relies on the kernel of the host OS. Sharing the kernel and avoiding the need for a guest OS typically yields better performance and a reduced footprint. One of the most popular implementations of containerization is Docker \cite{DockerDocumentationLocal}. Docker supports  process-level isolation, with two concepts from the kernel that are used: namespaces and cgroups. Namespaces are used to isolate certain resources: a network stack, its own processes tree and other resources. For example, a container cannot see or interact with the processes running in another container. The cgroups are used for specifying what computer resources a container should be able to access, such as certain amount of RAM, CPU cores, file system access and similar~\cite{Anderson2015}.

Docker containers can share namespaces. For example, sharing process ID (PID) namespace allows one container to access other containers' processes. Sharing an inter-process communication (IPC) namespace means that containers use the same network stacks and memory region. The feature of shareable namespaces in Docker is key for \chaosorca: it enables it to conduct experiments without modifying existing containers.

\subsection{Observability}

Observability~\cite{MilesRuss2019ChaosObservability} is the ability to collect values about some parts of the internal state of a system, based on its external behaviour. Increasing a system's observability is extremely valuable for debugging and reasoning about failures. Observability leverages three types of strategies: well designed event logs, sufficient monitoring metrics and tracing. An event log can be in different forms, plain text, structured or binary. Monitoring metrics usually combine information are different levels. For example, CPU and memory usage are useful for finding out problems like infinite loops or memory leaks. Also, the ratio of successful HTTP requests is helpful to describe whether the system is running under acceptable status. Tracing is a technique that combines distributed but related events together, such as a user request which spans multiple micro-services. Tracing is hard as it requires each application in the request path to propagate tracing information~\cite{Sridharan2018DistributedObservability}.

\subsection{Chaos Engineering}\label{sec:chaos-engineering}
Chaos engineering is a technique that evaluates resilience in production environment. When an application is deployed into production, it faces uncertainties like unstable network, unavailable third-party services and unbalanced traffic. A well designed application should be able to bear these unanticipated scenarios without any serious impact for the users. Tools for chaos engineering actively inject different kinds of failures into a production system and try to learn how the system behaves under perturbation. If the system behaves as expected, such experiments improve the confidence the developers have about the resilience of the system. Otherwise the experiment points out weaknesses in the ability of the system to handle failures~\cite{Basiri2016}.

There are five principles in chaos engineering that describe how developers should apply such a technique into their production system:
1) The very first one is to build a series of hypotheses about the system's steady state. The steady state is defined based on monitorable metrics. An hypothesis describes how the steady state is expected to change under a certain event which is triggered during chaos engineering experiments.
2) The second principle is that a failure injected by a chaos engineering experiment simulates a real-world event. For example, a simulated failure may be that the space of a disk is full, or a function returns with an exception.
3) As the goal of chaos engineering is to build confidence in a production system, chaos engineering experiments should be done after deployment, instead of in a testing environment.
4) Since it is impossible to predict every failure in a software system, chaos engineering experiments should be conducted continuously: automation is the fourth principle in chaos engineering.
5) The last but same important principle is ``blast radius control''. The term ``blast radius'' describes the seriousness of impacts caused by chaos engineering experiments. Running chaos engineering experiments are done so as to minimize negative influences on end-user experience.

\section{Design of \chaosorcabf}\label{sec:design}

\reviseadd{\chaosorca is addressing is an important problem: the lack of tools for developers to assess the resilience of containerized applications. \chaosorca contributes to addressing this problem in the qualified domain of evaluating resilience of containerized application with respect to system call errors. \chaosorca is designed to provide developers with observability and chaos engineering capabilities for an application container with respect to system call errors, without modification of the container. In other words, the core design principle is to gather sufficient insights about resilience, while causing minimal impact on the system under study. Thus, the requirements for designing \chaosorca are as follows:}

\begin{requirement}{Black-box containers:}\label{req:untouching}
\chaosorca should consider containers as black-boxes. In a production system, the images specifying the containers are often optimized for performance and security, meaning that it is impractical, if even possible, to weave-in extra instrumentation in the image directly. All action by \chaosorca should thus be implemented in a way that does not require instrumenting the containers. In other terms, \chaosorca should consider containers as pure, untouchable black-box entities.
A positive implication of this requirement is that it reduces the effort of developers to the minimum for using \chaosorca: an application container is deployed normally and \chaosorca is attached to it when developers need to run chaos experiments.
\end{requirement}

\begin{requirement}{Monitoring:}\label{req:monitoring}
As mentioned in~\autoref{sec:background}, observability is a foundation to analyze how the system behaviour differs during a chaos experiment. Given Requirement~\ref{req:untouching}, the internals of a container cannot be directly observed.  \chaosorca must monitor the surroundings of the containers and gather information to assess resilience weaknesses. This includes monitoring the impact that the containers have on the physical resources (RAM, CPU), on the I/O (network activity), and how they use system calls on the host kernel.
\end{requirement}

\begin{requirement}{System call perturbation:}\label{req:perturbation}
\chaosorca is designed for analyzing resilience with respect to system calls. Thus, \chaosorca should be able to perturb a specific type of system call. The form of perturbations also include 1) force a system call to directly return with an error code, 2) add a delay before a system call is actually completed, and 3) a combination of delay and error injection.
\end{requirement}

\begin{requirement}{Minimizing blast radius:}\label{req:blastradius}
In a production system, a server may host hundreds of containers, it is thus critical for \chaosorca to be able to target specific containers without influencing the others, so as to limit the blast radius of chaos experiments. Within this specific container, \chaosorca should be able to only perturb one single process as well.
\end{requirement}

\subsection{Overview of \chaosorcabf}
\begin{figure*}
    \centering
    \includegraphics[width=0.9\textwidth]{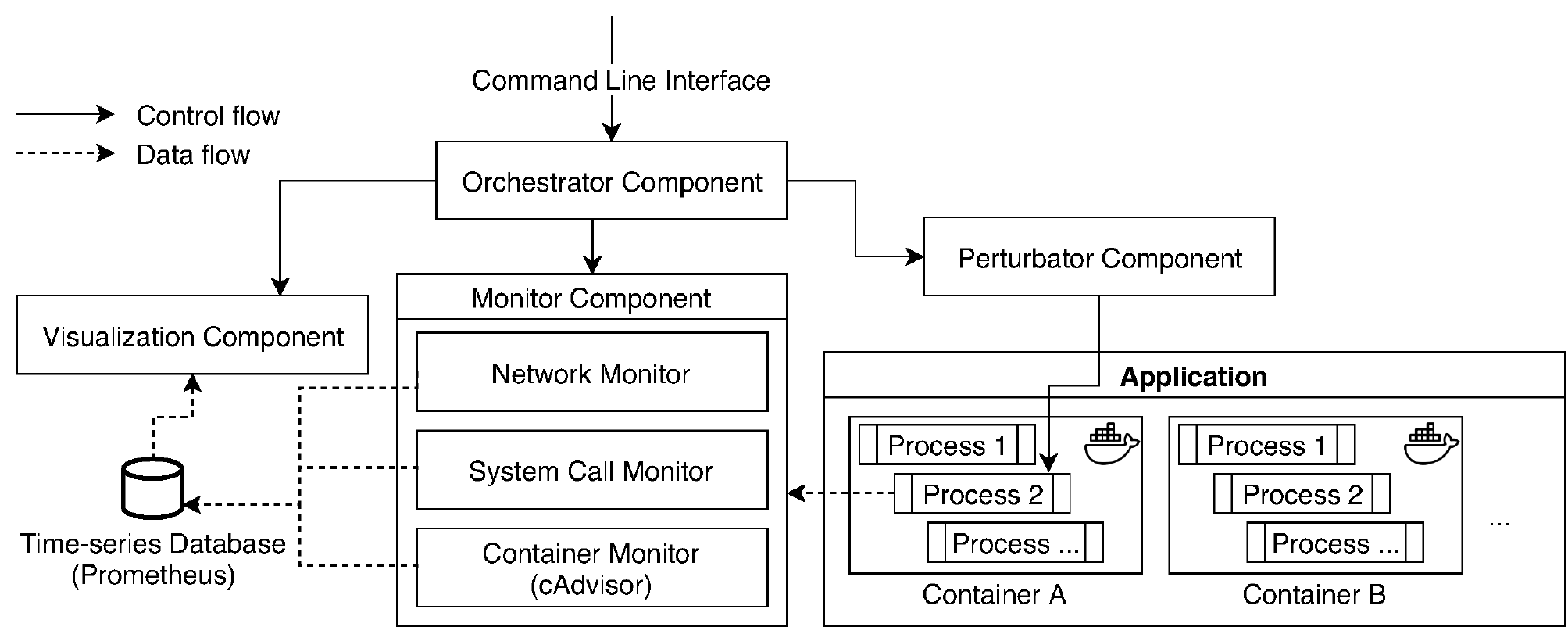}
    \caption{The Architecture of \chaosorca and The Interactions between Each Component: Solid Lines Stand for Control Flows. Dashed Lines Stand for Data Flows.}
    \label{fig:arch-over}
\end{figure*}

To fulfill the requirements listed above, \chaosorca is divided into three components: 1) monitor, 2) perturbator and 3) orchestrator. The interactions among them are described in~\autoref{fig:arch-over}. The monitor component is responsible for capturing the system behaviour at runtime. The perturbator injects different kinds of failures with respect to system calls. The orchestrator controls the monitor and perturbator components to conduct chaos experiments and generates reports.

\subsection{The \chaosorcabf Monitor Component}
Observability is essential to control and analyze chaos experiments. The monitor component implements requirement~\autoref{req:monitoring}. In order to connect low-level system call perturbations with application-level behaviour, the monitor component provides different levels of observability, respectively 1) container level, 2) system call level and 3) application level.

At the container level, the monitor records CPU time, memory usage, network traffic as follows. 
The monitoring component relies on shareable namespaces offered by Docker, it runs as a process which shares a specific namespace with the target container under study. This type of design fulfills requirement~\autoref{req:untouching}: the monitor provides strong observability with no modification or intrusion into the target container image.

For the system call level, the monitor component records the invoked system calls information including type, arguments and result. Regarding application level monitoring, the monitor cares about whether the application fulfills a user request, where the definition of fulfillment varies from one application to another. For example, for a web server, the monitor observes the HTTP response codes and the latency in handling HTTP requests.

\subsection{The \chaosorcabf Perturbator Component}\label{sec:perturbator}

The perturbator component is designed to inject system call failures at runtime. It addresses both requirement~\autoref{req:perturbation} and \autoref{req:untouching}. A system call perturbation is defined as a tuple $(s, e ,d)$ where $s$ represents the type of a system call, $e$ means the error code of this system call to be injected, and $d$ denotes the delay before this system call is actually invoked. 
In \chaosorca, the perturbator works as follows, it intercepts an existing system call that is invoked by the application, then ask for commands from the orchestrator (cf. \autoref{sec:orchestrator}) about whether injecting a specific failure in this system call. Per requirement~\autoref{req:perturbation}, the perturbator supports failures that could be either an error code, or a certain delay, or a combination of the both.

Let us consider the example of the TTorrent file download procedure shown in \autoref{fig:flamegraph}. When TTorrent needs to save a piece of downloaded file on disk, the logic in method \texttt{PieceStorageImpl/savePiece} is executed. This leads the JVM (i) to invoke a \texttt{pwrite} system call, and (ii) to write bytes of data into a file through this system call. Our perturbator can target such types of system calls, it can intercept this \texttt{pwrite} system call and injects a failure in TTorrent.

For the perturbator component, it is also important to minimize the blast radius, according to requirement~\autoref{req:blastradius}. Thanks to the shareable namespaces feature of Docker introduced in~\autoref{sec:bg-containerization}, \chaosorca is able to accurately inject system call invocation failures into a specific container only to minimize the blast radius. Regarding over-consumption issues an injected failure may cause, the blast radius can be still controlled by limiting the resources (CPU, RAM, etc.) a container is allowed to use.~\cite{dockerdocsRuntimeOptions}

\subsection{The \chaosorcabf Orchestrator Component}\label{sec:orchestrator}

The orchestrator component acts as an interface between \chaosorca and the fault injection strategies defined by the developers. It provides a command line interface for developers who want to apply \chaosorca. Based on the configuration, the orchestrator controls the monitor and perturbator components to explore a set of system call perturbations in the target application. Such a procedure is described in~\autoref{algo:core}. Line $1$ and $2$ are executed to gather the application's normal behaviour. The triple for-loops construct all possible system call failures and gather the perturbed behaviour for a comparison.

There are three phases for a fault injection experiment: 1) before perturbation, 2) during perturbation and 3) after perturbation as follows. First of all, before perturbation, the orchestrator attaches the monitor, the perturbator and itself to the target application. It feeds the application with a piece of workload in order to gather information about the normal behaviour of this application. After that, the orchestrator activates the perturbator to inject a specific system call failure defined in~\autoref{sec:perturbator}, this is the perturbation phase. Finally, the perturbator is turned off and the after-perturbation phase starts, the orchestrator keeps pulling monitoring information and computes the metric differences. With those three phases, \chaosorca enables itself to detect whether a system call failure has an instantaneous influence (in phase during-perturbation) or it has a long lasting influence even after the perturbation (in phase during-after).

Regarding the behaviour analysis in the during- and after- phases in our experiments, \chaosorca considers a mix of generic and domain specific dimensions.
The generic dimension is 1) whether the application continues to run without a crash and 2) how the metrics under monitoring vary compared to nominal behaviour. \chaosorca also considers domain specific dimensions. For example, for web applications, \chaosorca compares whether the HTTP response code is $200$. A procedure of comparing an application's perturbed behaviour with normal behaviour is called ``Diff'' in \autoref{algo:core}.

\begin{algorithm}[tb]
\caption{Fault Injection Experiments for Containers}
\label{algo:core}
\begin{algorithmic}[1]
\REQUIRE ~~\\ 
An application Docker container $A$;\\
A repeatable workload for this application $W$;\\
A set of system call types $T$;\\
A set of system call error codes $E$;\\
A set of delay values $D$;\\
\ENSURE ~~\\ 
$R$ a map of behaviour difference per perturbation;

\STATE Attach monitor, feed $A$ with $W$
\STATE normal\_behaviour $\leftarrow$ monitored metrics;
\FOR{each syscall type $t \in T$}
  \FOR{each error code $e \in E \cup \{0\}$}
    \FOR{each delay $d \in D \cup \{0\}$}
      \IF{$e == 0$ and $d == 0$}
        \STATE Continue; // same to normal execution
      \ENDIF
      \STATE perturbation $p \leftarrow <t, e, d>$;
      \STATE Feed $A$ with $W$ under $p$
      \STATE perturbed\_behaviour $\leftarrow$ monitored metrics;
      \STATE $diff \leftarrow$ Diff(normal\_behaviour, perturbed\_behaviour)
      \STATE $R \leftarrow R \cup \{<t, e, d>: diff\}$
    \ENDFOR
  \ENDFOR
\ENDFOR
\RETURN $R$;
\end{algorithmic}
\end{algorithm}

\subsection{The \chaosorcabf Visualization Component}\label{sec:visualization}

By default all metrics monitored by \chaosorca are saved into a time-series database. The developers can query this data as a table and investigate them after fault injection experiments. \chaosorca also has a separate component that visualizes all the monitored metrics as line charts.

For the convenience of resilience analysis, the following features are supported by the visualization component: 1) auto-refresh the charts in a configurable time rate, 2) highlight the perturbations, 3) aggregate the metrics with a developer defined configuration. For example, one chart can display the rate of all different system calls with a legend. In this way it helps developers to intuitively compare the rate among different system calls at a certain time point.

\begin{figure}
    \centering
    \includegraphics[width=.6\columnwidth]{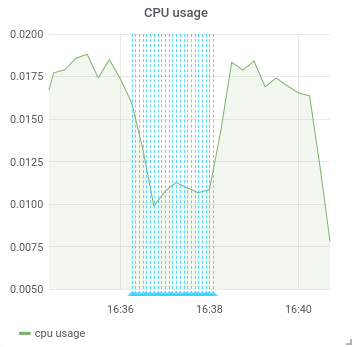}
    \caption{\chaosorca Visualization Component, with one monitored metric (green line), and one perturbation campaign (blue overlay)}
    \label{fig:visualization-design}
\end{figure}

\autoref{fig:visualization-design} shows an example of the visualization component. The line chart describes how the CPU usage changes during fault injection experiments. A blue overlay indicates a period of system call perturbations, which also divides the figure into three phases: before, during, and after the perturbations. The green line explains that the CPU usage drops during the perturbations and catches up afterward. The \chaosorca visualization component draws one line chart for each of the monitored metrics similarly.

\subsection{Implementation}\label{sec:implementation}
The monitor component uses cAdvisor\footnote{\url{https://github.com/google/cadvisor}} to gather container\hyp{}specific metrics like CPU usage, memory usage and network traffic metrics. For monitoring HTTP requests, the monitor combines PyShark\footnote{\url{https://kiminewt.github.io/pyshark/}}, a python library for TShark to get HTTP response code and latency. For collecting the system calls the monitor uses the open-source tool Bpftrace\footnote{\url{https://github.com/iovisor/bpftrace}} to observe the number and type of system calls. All the above monitoring information is pushed to a Prometheus time-series database\footnote{\url{https://prometheus.io/}}. 

Then \chaosorca uses Grafana\footnote{\url{https://grafana.com/}} to visualize the data for developers. The perturbator component is mainly implemented with Strace \footnote{\url{https://strace.io/}}, which is a diagnostic, debugging and instructional userspace utility for Linux. The orchestrator component itself is written in Python. It provides a command line interface to developers to define system call perturbation arguments like type, error code and delay. \chaosorca is publicly-available at \href{{https://github.com/KTH/royal- chaos/tree/master/chaosorca}}{https://github.com/KTH/royal-chaos/tree/master/chaosorca}.

\section{Evaluation}\label{sec:evaluation}
This section discusses the evaluation of \chaosorca, which focuses on the following 4 research questions:

\newcommand\rqsystemcall{What system call perturbations are of interest for containerized applications?
\xspace}
\begin{question}\label{rq1}
\rqsystemcall
\end{question}

\newcommand\rqproblemidentification{What kind of resilience problems are identified with fault injection in containerized applications?}
\begin{question}\label{rq2}
\rqproblemidentification
\end{question}

\newcommand\rqobservabilityimprovement{How is multi-layer observability helpful to analyze the results of chaos experiments?}
\begin{question}\label{rq3}
\rqobservabilityimprovement
\end{question}

\newcommand\rqoverhead{What is the overhead of conducting chaos experiments using \chaosorca?}
\begin{question}\label{rq4}
\rqoverhead
\end{question}

\subsection{Subject Programs}\label{sec:subject-programs}
We experiment \chaosorca on three different applications, in order to evaluate its ability at perturbing system calls and at detecting resilience weaknesses in containerized applications.
We selected the applications according to the following criteria:
1) medium-sized, well implemented applications with source code available, 2) contain both client side and server side code 3) can be deployed as Docker containers and 4) it is possible to simulate production-like workloads in the laboratory. 

The first application is TTorrent\footnote{\url{https://github.com/mpetazzoni/ttorrent}}, a file downloading client that uses the BitTorrent protocol. The second application is Nginx\footnote{\url{http://hg.nginx.org/nginx/file/tip}}, a popular and mature reverse proxy. The last one is Bookinfo\footnote{\url{https://github.com/istio/istio/tree/master/samples/bookinfo}}, a micro-service based reference application which runs several containers written in different languages. 
The key descriptive metrics of those subjects are shown in \autoref{tab:subject-programs} (e.g the number of containers, programming languages, lines of code). Note that \chaosorca focuses on a subset of containers in a larger software system and leave the other containers running normally. To that extent, it would likely work on perturbing one single isolated container in a very large system.

\begin{table}
\centering
\caption{Descriptive Statistics of The Subject Programs}\label{tab:subject-programs}
\scriptsize
\begin{tabular}{lp{2cm}p{0.8cm}ll}
\toprule
Application& Desc.& Lang.& LoC& \#Containers\\
\midrule
TTorrent& File downloading tool& Java& 14.3K& 1\\
Nginx& Reverse proxy& C& 141.1K& 1\\
Bookinfo& Micro-service based web application& Java, Python, Ruby& 10.0K& 6\\
\bottomrule
\end{tabular}
\end{table}

\subsection{Experimental Protocol}
\label{sec:protocolcs}
\textbf{Protocol for RQ1:}
In order to select the most interesting system calls for chaos experiments, we consider two dimensions: 1) the prevalence of a system call during execution 2) the functionality of the system call under study. The first dimension is investigated by monitoring applications. The second one needs more domain knowledge. For example, an application which frequently writes files invokes $write$ a lot. So $write$ takes a large ratio in all happened system calls. Before invoking $write$, another system call \texttt{open} needs to be invoked. It opens the target file and returns the file descriptor. \texttt{open} system call may not be invoked lots of times, but it is still interesting since $write$, $read$ system calls rely on it.

Every application runs for $1$ minute under a simulated workload, and we perform $3$ rounds of executions in total for a statistical reason. In this experiment, \chaosorca only monitors the applications and does not inject any failure. The workloads are as follows: TTorrent is executed to download an Ubuntu installation file from the internet, the file is 1.86GB in size. For Nginx, three types of valid GET requests including \texttt{/}, \texttt{/abc} and \texttt{/abc?cache=bust} are sent to the container. For Bookinfo, a collection of valid GET and POST requests are used to trigger the following features: the login logic; accessing the main page; accessing a product page and logging out. 
For all three applications, \chaosorca calculates the rate of different types of system calls.

\textbf{Protocol for RQ2:}
For each system call identified as relevant in RQ\autoref{rq1}, three different perturbation strategies are evaluated according to~\autoref{algo:core}: 1) injecting a system call error code without any delay 2) injecting a certain delay after which the system call succeeds and 3) injecting both an error code and a delay. Considering there are more than $100$ different error codes that can be returned by system calls~\cite{LinuxSystemCalls}, we carefully select $6$ error codes such that they are all related to resource and permission issues~\cite{GNU-libc}, namely \texttt{EACCES}, \texttt{EPERM}, \texttt{ENOENT}, \texttt{EIO}, \texttt{EINTR} and \texttt{ENOSYS}. We do so because these system call errors are representative of real-life failures. For example, error code \texttt{EACCES} means ``permission denied'', indicating that the file permissions do not allow the attempted operation. Then we run chaos experiments with these $6$ error codes and $2$ different delay values ($1$s and $5$s). The workload for these applications are the same with RQ\autoref{rq1}. Both generic and domain specific dimensions described in \autoref{sec:orchestrator} are used for behaviour comparison.

\textbf{Protocol for RQ3:}
The experiment protocol is the same as for RQ2. However, for this research question, we focus on monitoring and data visualization. Consequently, RQ3 is discussed qualitatively. For example, by removing part of observability provided by \chaosorca, such as closing HTTP response code monitoring, we analyze whether it is still possible to correctly detect resilience problems.

\textbf{Protocol for RQ4:}
A study of the overhead of \chaosorca is important to assess its applicability. To answer RQ4, we respectively evaluate the overhead caused by the monitor component and the perturbator component. For the network overhead, Nginx is used as a target application. Then, the tool Siege~\footnote{Siege (https://github.com/JoeDog/siege) is an http load tester which generates requests at a fast rate} is used to simulate a number of concurrent users sending requests for $3$ minutes. The number of users is respectively $5, 10, 25, 200, 1000, 2000$ during the experiment.

For system call monitoring and perturbation overhead, in order to stimulate a high amount of system calls, \texttt{dd} command is used as a target application. \texttt{dd} is a Linux command which mainly coverts and copies files. It invokes a huge amount of $read$ and $write$ system calls. The command to test \texttt{dd} is \texttt{dd if=/dev/zero of=/dev/null bs=1 count=500000k}, which copies zero to the void for 500 million times.

\subsection{RQ1: \rqsystemcall}

The statistics of the invoked system calls are listed in~\autoref{tab:system-call-statistics}. Each line describes one of the system call invoked during the execution. Each column shows the number and percentage of system call invocations in one containerized application. A blank cell means that a system call is never invoked by the corresponding application under the given workload. Some system calls are rarely used by the application, which are noted as $< 0.01$. For example, the first row of the table tells that TTorrent invokes the \texttt{accept} system call $1780$ times every second, which takes about $9.92\%$ of the total. While Nginx never invokes the $accept$ system call, Bookinfo invokes it $293.45$ times per second which takes $3.90\%$ of the total.

\begin{table}
\centering
\caption{The Statistics of System Calls per Second for Each Application under The Given Workload without Any Perturbations}\label{tab:system-call-statistics}
\scriptsize
\begin{tabular}{lllll}
\toprule
System Call& TTorrent& Nginx& Bookinfo\\
\midrule
accept& 1780 (9.92\%)& & 293.45 (3.90\%)\\
accept4& & 21.21 (1.17\%) & \\
bind& & & 173.57 (2.31\%)\\
clone& & 166.57 (9.17\%)& 232.77 (3.09\%)\\
close& 92 (0.51\%)& & 280.37 (3.73\%)\\
connect& & & 163.03 (2.17\%)\\
ctl& & 19.24 (1.06\%) & \\
dup& & & 175.95 (2.34\%)\\
dup2& 64.31 (0.36\%)& & \\
exit& 62.4 (0.35\%)& & 174.36 (2.32\%)\\
fcntl& & & 161.89 (2.15\%)\\
futex& 3240 (18.06\%)& & 854.96 (11.36\%)\\
getpid& & 160.52 (8.84\%)& 159.6 (2.12\%)\\
getsockname& & 48.77 (2.69\%) & 174.84 (2.32\%)\\
getsockopt& & & 167.27 (2.22\%)\\ 
ioctl& & & 189.22 (2.51\%)\\
list& & & 247.43 (3.29\%)\\
lseek& < 0.01& < 0.01& \\
madvise& 79.28 (0.44\%)& & 176.86 (2.35\%)\\
mmap& 191.09 (1.07\%)& & < 0.01\\
mprotect& 677 (3.77\%)& & 15.74 (0.21\%)\\
munmap& 0.03 (< 0.01\%)& & < 0.01\\
newfstat& < 0.01& 140.56 (7.74\%) & 206.17 (2.74\%)\\
newlstat& & & \\
newstat& & 89.96 (4.95\%)& 1300 (17.27\%)\\
newuname& & & \\
open& < 0.01& 166.15 (9.15\%) & 222.19 (2.95\%)\\
poll& & & 181.12 (2.41\%)\\
pread64& 1410 (7.86\%)& 99.37 (5.47\%) & \\
pwrite64& 1520 (8.47\%)& 147.4 (8.12\%) & \\
read& 3570 (19.90\%)& & 158.48 (2.11\%)\\
recvfrom& & 90.37 (4.98\%) & 130.69 (1.74\%)\\
recvmsg& & & 174.7 (2.32\%)\\
select& & & 388.57 (5.16\%)\\
sendfile64& & 264.61 (14.57\%) & \\
sendto& & & 161.78 (2.15\%)\\
setsockopt& & 29.41 (1.62\%) & 287.41 (3.82\%)\\
shutdown& & & 170.71 (2.27\%)\\
sigprocmask& 46.78 (0.26\%)& & \\
socket& & & 263.94 (3.51\%)\\
wait& 3040 (16.95\%)& & \\
write& 2110 (11.76\%)& 87.04 (4.79\%)& 238.31 (3.17\%)\\
writev& & 284.63 (15.68\%) & \\
yield& 55.18 (0.31\%)& & \\
\bottomrule
\end{tabular}
\end{table}

From~\autoref{tab:system-call-statistics} we could see that the types of most used system calls differ in applications. For example, TTorrent invokes \texttt{read} and \texttt{futex} most frequently. This actually makes sense because TTorrent takes advantages of multi-threading to download files. System calls like \texttt{read}, \texttt{futex} are keys to support TTorrent's functionality. Thus these most used system calls are of interest for evaluating TTorrent. Bookinfo invokes the most types of system calls during an experiment, because it is micro-service based and has a more complex business logic. For HTTP based applications including Nginx and Bookinfo, the system calls used in common are \texttt{clone}, \texttt{getpid}, \texttt{getsockname}, \texttt{newfstat}, \texttt{newstat}, \texttt{open}, \texttt{recvfrom}, \texttt{setsockopt} and \texttt{write}.

With the help of \chaosorca, developers are able to focus more on the most critical system calls per application. This makes the following resilience evaluation more targeted and effective.

\begin{mdframed}[style=mpdframe,nobreak=true,frametitle=Answer to RQ1]
Over our three case studies, we observe a total of $44$ different types of system calls exercised by the containerized applications under study.
By analyzing the prevalence and the functionality of each of them, $9$ types of system calls are identified as relevant for fault injection experiments in containerized applications, namely \texttt{open}, \texttt{write}, \texttt{writev}, \texttt{read}, \texttt{readv}, \texttt{sendfile}, \texttt{sendfile64}, \texttt{poll} and \texttt{select}.
\end{mdframed}

\subsection{RQ2: \rqproblemidentification}

According to the experiment protocol, we have conducted $9x(6+2+6x2)=180$ rounds of experiments. After finishing all the experiments, we qualitatively analyzed the influence caused by each perturbation. In the following, we discuss the most used system calls. The interested reader can access to the comprehensive experimental data at \href{https://github.com/KTH/royal-chaos/tree/master/chaosorca/experiment\_data}{https://github.com/K TH/royal-chaos/tree/master/chaosorca}.

\subsubsection{Evaluations on TTorrent}

\autoref{tab:summary-ttorrent} shows a sample of experiment results in TTorrent. As it is a file downloading client, system calls like \texttt{open}, \texttt{write}, \texttt{read} are the most used ones. In this table, each row describes one perturbation and TTorrent's corresponding behaviour. Each cell is the result of the comparison of the application's behaviour under and after perturbation.

\begin{table*}[pos=t]
\centering
\caption{Representative Sample of Chaos Experiments on TTorrent}\label{tab:summary-ttorrent}
\scriptsize
\begin{tabular}{@{\makebox[3em][r]{\stepcounter{rowcount}\therowcount\hspace*{\tabcolsep}}}lllllllll}
\toprule
System Call& Error Code& Delay& Trend of Syscalls& Network IO& Cpu Usage& Memory Usage\\
\midrule
open& -& 1s& fewer, spike& small dip& increase& normal\\
open& -& 5s& increase& dip& dip, increase& normal\\
open& EACESS& -& increase& normal& increase& normal\\
open& EACESS& 1s& increase& small dip& increase& normal\\
open& EACESS& 5s& increase& dip& dip, increase& normal\\
write& -& 1s& tiny& tiny& dip& normal\\
write& -& 5s& small& tiny& dip& normal\\
write& EACESS& -& tiny& stops& tiny, tiny& normal\\
write& EACESS& 1s& tiny, tiny spike& stops& tiny, tiny& normal\\
write& EACESS& 5s& tiny, small spike& stops& tiny, tiny& normal\\
write& EPERM& -& tiny& stops& tiny, tiny& normal\\
write& EPERM& 1s& tiny, small spike& stops& tiny, tiny& normal\\
write& EPERM& 5s& small, increase& tiny, small& tiny, small& normal\\
write& ENOENT& -& small, spike& stops& tiny, tiny& normal\\
write& ENOENT& 1s& spike, tiny& stops& tiny, tiny& normal\\
write& ENOENT& 5s& tiny, small spike& stops& tiny, tiny& normal\\
write& EIO& -& small, small spike& stops& tiny, tiny& normal\\
write& EIO& 1s& small, small spike& stops& tiny, tiny& normal\\
write& EIO& 5s& small, small spike& stops& tiny, tiny& normal\\
read& -& 1s& big dip& big dip& big dip& normal\\
read& -& 5s& big dip& big dip& big dip& normal\\
read& EACESS& -& tiny& stops& tiny, tiny& normal\\
read& EACESS& 1s& tiny& stops& tiny, tiny& normal\\
read& EACESS& 5s& small, small& stops& tiny, tiny& normal\\
read& EPERM& -& tiny& stops& tiny, tiny& normal\\
read& EPERM& 1s& tiny& stops& tiny, tiny& normal\\
read& EPERM& 5s& tiny& stops& tiny, tiny& normal\\
read& ENOENT& -& tiny& stops& tiny, tiny& normal\\
read& ENOENT& 1s& tiny& stops& tiny, tiny& normal\\
read& ENOENT& 5s& small& stops& tiny, tiny& normal\\
\bottomrule
\end{tabular}
\vspace{-0.3cm}
\end{table*}

From \autoref{tab:summary-ttorrent} it can be seen that only adding a delay to \texttt{open}, \texttt{write}, \texttt{read} system calls have a critical influence on network usage. Especially row $21$ and $22$ show that the $read$ system call perturbation causes the biggest dip in network IO. Recalling~\autoref{tab:system-call-statistics}, this is consistent because the \texttt{read} system call is the most common one, it occurs almost $20$ percent of the total number of calls.

Another insight from the table is that for the \texttt{open} system call perturbations, network IO dips initially during the perturbation and then catches up after. For \texttt{write} and \texttt{read} system call perturbations, the network traffic stops completely and the number of system calls drops to nearly zero. This indicates that TTorrent is able to bear system call delays but is more sensitive to system call errors.

\subsubsection{Evaluations on Nginx}

\autoref{tab:summary-nginx} describes the most interesting findings from our experiments on Nginx. For Nginx, the perturbations that have an obvious effect are those on \texttt{open}, \texttt{write}, \texttt{writev} and \texttt{sendfile}. As an example, the first row records the Nginx behaviour under an \texttt{open} system call perturbation: adding a $1$ second delay with no system call error code. However, the HTTP response code becomes $0$ and the latency increases to more than $50$ ms. The number of total system calls decreases during the perturbation compared to the normal execution. The network and CPU usage stay low while some spikes occur. The usage of memory increases during the experiment. This proves that Nginx is able to handle \texttt{open} system call delays and the usage of calculation resources stays acceptable.

\begin{table*}[pos=t]
\centering
\caption{Representative  Sample of Chaos Experiments on Nginx}\label{tab:summary-nginx}
\scriptsize
\setcounter{rowcount}{-1}
\begin{tabular}{@{\makebox[3em][r]{\stepcounter{rowcount}\therowcount\hspace*{\tabcolsep}}}llllllllll}
\toprule
System Call& Error Code& Delay& HTTP Res.& Latency& Trend of Syscalls& Network IO& Cpu Usage& Memory Usage\\
\midrule
open& -& 1s& 0& up after 50ms& fewer during& tiny, spike& tiny, spike& increasing\\
open& -& 5s& 0& up after 96ms& FEWER, spike& tiny, spike& tiny, spike& increasing\\
open& EACCES& -& 403& 0& fewer& small& small& increasing\\
open& EACCES& 1s& 403 (fewer)& increasing 102ms& fewer, spike& tiny, spike& tiny, spike& increasing\\
open& EACCES& 5s& 403 (FEW)& increasing 140ms& FEWER, spike& tiny, spike& tiny, spike& increasing\\
open& EPERM& -& 500& 0& fewer& small& smaller& increasing\\
open& EPERM& 1s& 500 (fewer)& increasing 57ms& normal, spike& tiny, spike& tiny, spike& increasing\\
open& EPERM& 5s& 500 (FEW)& increasing 175ms& FEWER, spike& tiny, spike& tiny, spike& increasing\\
open& ENOENT& -& 404& 0& fewer& small& small& increasing\\
open& ENOENT& 1s& 404 (fewer)& increasing 150ms& normal, spike& tiny, spike& tiny, spike& increasing\\
open& ENOENT& 5s& 404 (FEW)& increasing 97ms& FEWER, spike& tiny, spike& tiny, spike& increasing\\
open& EIO& -& 500& 0& fewer& small& smaller& increasing\\
open& EIO& 1s& 500 (fewer)& increasing 96ms& increase, spike& tiny, spike& tiny, spike& increasing\\
open& EIO& 5s& 500 (FEW)& increasing 172ms& FEWER, spike& tiny, spike& tiny, spike& increasing\\
open& EINTR& -& 500& 0& fewer& small& smaller& increasing\\
open& EINTR& 1s& 500 (fewer)& increasing 96ms& fewer, spike& tiny, spike& tiny, spike& increasing\\
open& EINTR& 5s& 500 (FEW)& increasing 70ms& FEWER, spike& tiny, spike& tiny, spike& increasing\\
open& ENOSYS& -& 500& 0& normal& small& smaller& increasing\\
open& ENOSYS& 1s& 500 (fewer)& increasing 149ms& normal, spike& tiny, spike& tiny, spike& increasing\\
open& ENOSYS& 5s& 500 (FEW)& increasing 159ms& FEWER, spike& tiny, spike& tiny, spike& increasing\\
write& -& 1s& 0& up after 75ms& fewer, spike& tiny, spike& tiny, spike& increasing\\
write& -& 5s& 0& up after 51ms& fewer, spike& tiny, spike& tiny, spike& increasing\\
write& EACCES& -& 200& 0& normal& normal& increase& increasing\\
write& EACCES& 1s& 200 (FEW)& up after 56ms& fewer, spike& tiny, spike& tiny, spike& increasing\\
write& EACCES& 5s& 0& up after 107ms& FEWER, spike& tiny, spike& tiny, spike& increasing\\
read& -& 1s& 200& 0& normal& normal& increase& increasing\\
read& -& 5s& 200& 0& normal& normal& increase& increasing\\
read& EACCES& -& 200& 0& normal& normal& increase& increasing\\
read& EACCES& 1s& 200& 0& normal& normal& increase& increasing\\
read& EACCES& 5s& 200& 0& normal& normal& increase& increasing\\
sendfile& -& 1s& 0& up after 35ms& fewer, spike& tiny, spike& tiny, spike& increasing\\
sendfile& -& 5s& 0& up after 172ms& FEWER, spike& tiny, spike& tiny, spike& increasing\\
sendfile& EACCES& -& 0& 0& increase& small& normal& increasing\\
sendfile& EACCES& 1s& 0& up after 59ms& normal, spike& tiny, spike& tiny, spike& increasing\\
sendfile& EPERM& -& 0& 0& normal& small& normal& increasing\\
sendfile& EPERM& 1s& 0& up after 64ms& fewer, spike& tiny, spike& tiny, spike& increasing\\
\bottomrule
\end{tabular}
\vspace{-0.3cm}
\end{table*}

Row $1$ to row $20$ shows that an \texttt{open} system call perturbation has different effects depending on the injected error code. An $EACCES$ error results in HTTP 403 Forbidden, an $ENOENT$ error leads to 404 Not Found instead. An $EPERM$, $EINTR$, or $ENOSYS$ results in HTTP 500 Internal Server Error. This could be considered as a good practice because developers designed a more detailed reaction strategies with respect to different system call errors.

Another interesting insight to look at is the \texttt{read} system call perturbations (row $26$ to row $30$). When such a perturbation is injected, the application still functions normally as the HTTP response code is still $200$. The trend of system calls and the network IO stay normal as well. According to these experiments, developers gain more confidence that Nginx is able to survive from \texttt{read} system call errors.

\subsubsection{Evaluations on Bookinfo}

\begin{table*}[pos=t]
\centering
\caption{Representative Sample of Chaos Experiments on Bookinfo}\label{tab:summary-bookinfo}
\scriptsize
\begin{tabular}{llllllllll}
\toprule
System Call& Error Code& Delay& HTTP Res.& Latency& Trend of Syscalls& Network IO& Cpu Usage& Memory Usage\\
\midrule
open& -& 1s& 200 / 302& increase 5ms& normal& normal& increase& normal\\
open& -& 5s& 200(lower) / 302& up after 2ms& increase& normal& increase& increase\\
open& EACESS& -& 200(lower) / 302& 0& fewer, up& smaller& increase& normal\\
read& -& 1s& 200(fewer) / 302& increasing 2ms& increase& smaller& increase& increase\\
read& -& 5s& 200(FEW) / 302& up after 4ms& increase, spike& small, spike& increase, spike& increase\\
read& EACESS& -& 0& 0& normal& small& smaller& normal\\
read& EACESS& 1s& 0& up after 1ms& increase, spike& small, spike& smaller, spike& increase\\
read& EACESS& 5s& 0& up after 1ms& normal, spike& small, spike& smaller, spike& increase\\
select& -& 1s& 200(FEW) / 302& up after 6ms& normal, spike& tiny, spike& smaller, spike& increase\\
select& -& 5s& 0& up after 12ms& tiny, spike& tiny, spike& small, spike& increase\\
select& EACESS& -& crash& -& -& -& -& -\\
select& EACESS& 1s& crash& -& -& -& -& -\\
select& EACESS& 5s& crash& -& -& -& -& -\\
select& EPERM& -& crash& -& -& -& -& -\\
select& EPERM& 1s& crash& -& -& -& -& -\\
select& EPERM& 5s& crash& -& -& -& -& -\\
\bottomrule
\end{tabular}
\vspace{-0.3cm}
\end{table*}

\autoref{tab:summary-bookinfo} gives a representative sample of our experimental results. For the Bookinfo application, the system call perturbations that have a significant effect are \texttt{open}, \texttt{write}, \texttt{read}, \texttt{poll} and \texttt{select}. Compared to Nginx, when \chaosorca injects delays into the \texttt{open} system call, the code of responses are divided into $200$ and $302$. Code $302$ means a redirection to another url. Considering the Bookinfo containers as a black box, we assume that this is because the injected delay causes a timeout exception in Bookinfo's components. In order to lead users to a retry logic, Bookinfo redirects the request to an initial page. For the \texttt{select} system call perturbations, Bookinfo directly crashes if an error code is injected. This indicates that the application is sensitive to \texttt{select} system call errors, and its self-protection strategies are not able to pull Bookinfo back to normal when such a perturbation happens.

\begin{mdframed}[style=mpdframe,nobreak=true,frametitle=Answer to RQ2]
By performing fault injection in three different containerized applications, we show that two kinds of resilience problems are identified by \chaosorca: 1) the application is not resilient to some system call failures: if certain system calls fail, the whole application crashes and 2) the application is sensitive to system call errors: if certain system calls fail, the application consumes significantly more system resources.
\end{mdframed}

\subsection{RQ3: \rqobservabilityimprovement}

In order to improve the observability of target applications, the monitoring component in \chaosorca records different levels of metrics: 1) system call level, 2) container level and 3) application level. In order to easily compare the changes under perturbation with a normal execution, all of these metrics are visualized by \chaosorca in a Grafana dashboard.

\begin{figure}
    \centering
    \subfloat[Network IO]{{\includegraphics[width=.45\columnwidth]{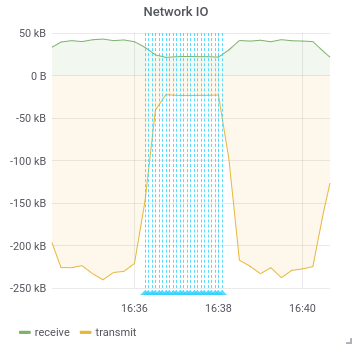}}}%
    \subfloat[HTTP Latency]{{\includegraphics[width=.45\columnwidth]{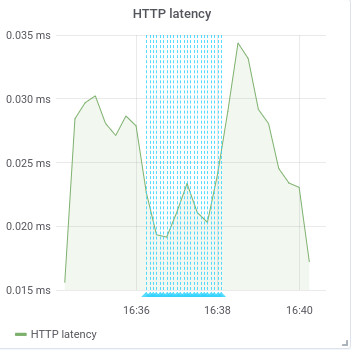} }}%
    \\
    \subfloat[Memory Usage]{{\includegraphics[width=.45\columnwidth]{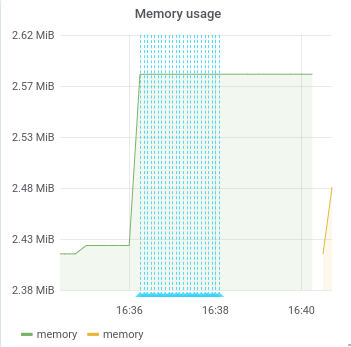} }}
    \subfloat[CPU Usage]{{\includegraphics[width=.45\columnwidth]{images/visualization/nginx-open-eacess-cpu.png} }}
    \\
    \subfloat[Rate of System Calls]{{\includegraphics[width=.45\columnwidth]{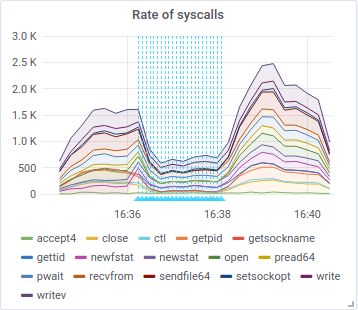} }}%
    \subfloat[Rate of HTTP request and response code]{{\includegraphics[width=.45\columnwidth]{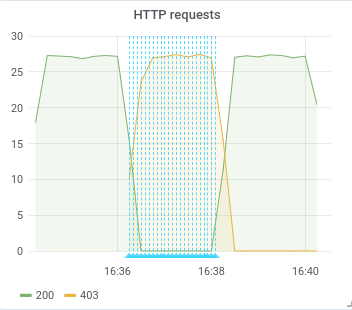} }}%
    \caption{\chaosorca Visualization: Nginx with OPEN system call EACCES error perturbation. A blue overlay shows when an perturbation is happening.}%
    \label{fig:visualization-example}%
    \vspace{-0.5cm}
\end{figure}

For example, \autoref{fig:visualization-example} is a metrics visualization of an experiment on Nginx. In this series of experiments, \chaosorca injects an \texttt{EACCES} error every time when the \texttt{open} system call is invoked. In the figure, a blue line stands for a specific system call perturbation. From the figure it can be seen that \chaosorca monitors Nginx before any perturbations for $5$ minutes. Then it keeps intercepting \texttt{open} system calls and injecting an \texttt{EACCES} error in them. After the perturbator is turned off, \chaosorca monitors Nginx for another $5$ minutes. Thanks to the monitor component, the $5$ metrics collected by \chaosorca namely network IO, HTTP latency, CPU and memory usage, the amount of invoked system calls are observable by developers. The visualization also makes it easier for developers to analyze the three experiment phases defined in \autoref{sec:orchestrator}: the trend of metric changes before, during and after perturbations.

On the other hand, if the observability of target application is not sufficient, it brings more obstacles to distinguish abnormal behaviour. For example, during an \texttt{OPEN} system call perturbation in \autoref{fig:visualization-example}, the rate of HTTP requests stays the same. However, most of requests are responded with a $403$ ``Forbidden'' code according to sub-figure (f). If \chaosorca does not bring observability to web application's HTTP response code, developers may think that everything is fine since Nginx still handles a proper amount of requests. But actually most of the responses are incorrect. Such a perturbation does have critical influence on end-users.

\begin{mdframed}[style=mpdframe,frametitle=Answer to RQ3]
For a containerized applications, there are different locations in the stack that enable engineers to improve observability.
\chaosorca shows the feasibility of multi-layer monitoring at the system call level, the container level and the application level.
\end{mdframed}

\subsection{RQ4: \rqoverhead}

\subsubsection{The Overhead of the Monitoring Component}

\begin{figure}
    \centering
    \subfloat[HTTP requests]{{\includegraphics[width=.45\columnwidth]{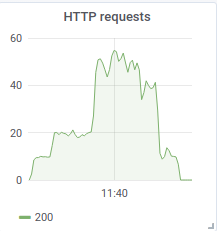} }}
    \subfloat[System calls]{{\includegraphics[width=.45\columnwidth]{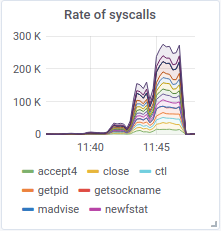} }}
    \newline
    \subfloat[CPU Usage]{{\includegraphics[width=.45\columnwidth]{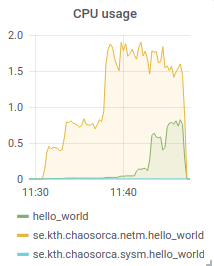}}}
    \subfloat[Memory Usage]{{\includegraphics[width=.45\columnwidth]{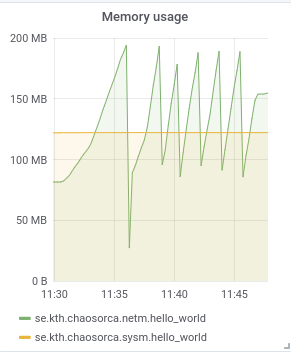}}}
    \caption{{The Overhead of Monitor Component when Nginx Is The Target Application}}
    \label{fig:overhead-monitor}
    \vspace{-0.5cm}
\end{figure}

\autoref{fig:overhead-monitor} shows the overhead of monitor component when Nginx is the target application. The first sub-figure records the amount of HTTP requests per second during the experiment. The second sub-figure records the rate of system calls. The last two sub-figures are about the CPU and memory usage of the monitor components. Regarding \texttt{dd} command, the CPU and memory usage metrics are the same as Nginx's ones.

For network monitoring, with the increase of HTTP request rate, sub-figure (a) in \autoref{fig:overhead-monitor} shows that the HTTP request monitoring even fails to monitor every request when the workload becomes too high. Sub-figure (c) describes that the CPU usage of network monitoring increases to almost two times of the application's itself. The memory usage cost by network monitor ranges from 50MB to 200MB which is shown in sub-figure (d).

For system call monitoring, the overhead is quite low and stable. Even under a system call heavy tool like \texttt{dd}, the CPU usage of the system call monitor module (chaosorca.sysm) is almost zero (sub-figure c). The memory usage stays at 125MB and never increases (sub-figure d).

Overall for monitor component, the overhead of network monitoring becomes significant when the rate of user requests are higher than $50/s$ in the test environment. The overhead caused by system call monitoring always stays low. As the memory usage, it is stable and there is no memory leak during the experiment. The overhead of \chaosorca can be considered as acceptable for a normal amount of workload.

\subsubsection{The Overhead of Perturbator Component}

\begin{figure}
    \centering
    \subfloat[HTTP Requests]{{\includegraphics[width=.45\columnwidth]{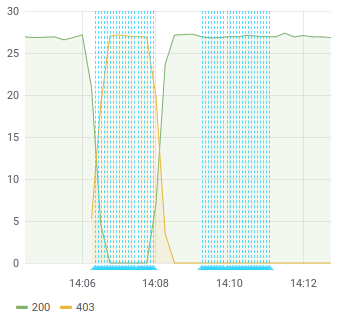}}}
    \subfloat[Rate of System Calls]{{\includegraphics[width=.45\columnwidth]{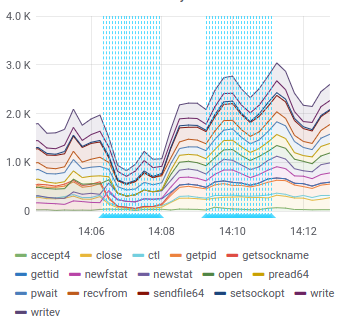}}}
    \caption{{Monitoring during A Perturbation. Each Blue Line Is A Perturbation. The First Group Affects Nginx's Behaviour And The Second One Does Not.}}
    \label{fig:overhead-perturbator-nginx}
    \vspace{-0.5cm}
\end{figure}

\begin{figure}
    \centering
    \includegraphics[width=.6\columnwidth]{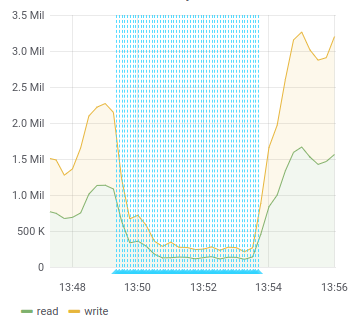}
    \caption{The Rate of System Calls Invoked By \texttt{dd} during A Perturbation. Each Blue Line Is When A Perturbation Occurs. }
    \label{fig:overhead-perturbator-dd}
\end{figure}

When the perturbator injects system call failures, the key metrics are described in \autoref{fig:overhead-perturbator-nginx} and \autoref{fig:overhead-perturbator-dd}. For Nginx, it shows that the number of HTTP requests under a normal workload remains the same. For the system call heavy application \texttt{dd}, the perturbations have an significant impact on the number of system calls. Overall for the perturbator component, the usage of system resources remains small during experiments.

\begin{mdframed}[style=mpdframe,frametitle=Answer to RQ4]
The overhead caused by \chaosorca depends on system call usage and also on the workload. For the monitor component, the network monitor overhead increases with the workload; and the system call monitor overhead is low. 
As for the perturbator component, our experiment shows that the overhead is acceptable. Overall, our experiments suggest that the runtime overhead of \chaosorca is acceptable for analyzing resilience of containerized applications.
\end{mdframed}

\section{Discussion}\label{sec:discussion}

\subsection{Threats to Validity}

One threat to the validity of \chaosorca is the perturbation search space has not been completely explored. We manually selected the $6$ most general and relevant system call error codes for our experiments. However, there are more than $100$ error codes that can be returned by different system calls. The interactions between a specific system call and its possible error codes have not been exhaustively analyzed. Future work will analyze the perturbation space with respect to all system call error codes.

Another threat is the generalization of the experimental results. \chaosorca has been evaluated on three different types of applications, including a file downloading client written in Java, a reverse proxy server written in C and a polyglot micro-service based application which contains 6 containerized services. Though these applications are selected according to the criteria mentioned in \autoref{sec:subject-programs}, conducting more experiments on other kinds of applications may change some answers to the research questions. Future work is required to explore other application domains and evaluate the generalization of the experimental results in \autoref{sec:evaluation}.

Lastly, in order to exclude accidental effects, we make $3$ rounds of experiments for each application (see \autoref{sec:protocolcs}). This is not enough for statistical analysis. It is an interesting area of future work to study statistically causal impact analysis on the fault injection results, which is little researched area.

\subsection{Extensibility of \chaosorcabf}
As introduced in \autoref{sec:implementation}, \chaosorca is implemented using specific features like namespace sharing in Docker. This limits \chaosorca to be applied by only Docker applications. However, the design of \chaosorca can be reused for other container systems like Singularity\footnote{\url{https://sylabs.io/}}. Furthermore, Singularity allows to interface with all of the resources, devices and network inside the container from the outside of the container~\cite{Singularity_User_Guide}. Theoretically, by tuning the commands used by \chaosorca's orchestrator component and applying specific monitoring tools for Singularity, \chaosorca can be used for the Singularity container system. Another well-designed container system is Shifter\footnote{\url{https://github.com/NERSC/shifter}}, which is particularly used for high performance computing. However, Shifter does not use network or process namespaces, nor cgroups. The implementation of ChaosOrca can not be easily adapted for Shifter. Future work is needed to evaluate the extensibility and overhead of \chaosorca in other container systems.

\subsection{Resilience Observation With or Without an Application Container}

ChaosOrca aims at allowing observability and fault injection in containers, in order to assess resilience. All of the research questions and the corresponding experiments are designed based on a the presence of a container environment. It is assumed that the resilience observed inside and outside the container is similar. Future work is needed to systematically research the difference between the resilience observed inside and outside an application container, i.e. whether containerization itself is responsible for changing certain aspects of resilience.

\section{Related Work}\label{sec:relatedwork}

\subsection{Fault Injection}

Fault injection was originally studied at the hardware level, for example using radiations to force bit flips~\cite{gunneflo1989evaluation} or by interacting with pins to generate processor errors~\cite{madeira1994rifle}. Later on, software-based approach were developed to emulate hardware errors~\cite{kanawati1992ferrari} e.g. by injecting memory or network errors~\cite{han1993doctor} or by perturbing the OS kernel~\cite{KaoW.-I1993FAfi,HyosoonLee2000Sasi}. A number of tools now facilitates the injection of such kinds of errors, such as Gremlin~\cite{GremlinLocal}, ChaosCat~\cite{chaoscatLocal} or cpu-troll~\cite{cpu-trollLocal}. The network layer is another fault injection vector, which has been studied in depth. Quite a few open-source tools~\cite{Izrailevsky2011TheArmyLocal,AlexeiLedenev2017,Blockade,muxyLocal,comcastLocal} and closed-source tools~\cite{GremlinLocal,chaoscatLocal} combine Iptables with the Traffic Control network emulation tool to inject different kinds of network failures, including latency and dropping a percentage of traffic.

More related to our work, a number of approaches propose to inject faults at the system call level, for example by introducing bit flip during the execution of system calls~\cite{amarnath2018fault}, by corrupting those system calls~\cite{10.1007/3-540-48254-7_11}, or by overriding parts of the glibc library~\cite{LeeHyeong-Chan2011Ewsa}, which wraps system calls into a C API. Besides, approaches have been developed to better control and protect the access to those system calls~\cite{ThuYeinWin2017PCPI}. Containerization, cgroups and namespaces, which we presented in Section~\ref{sec:background}, provide process-level isolation and allows to finely control which resources a given container can access. \reviseadd{Yet the existing works above need to modify operating systems or core libraries, thus they are not applicable for a production environment. On the contrary, since \chaosorca requires no modification at all and has a low overhead, it is applicable in production.}

Chaos Engineering is an emerging discipline, building on those fault injection tools, but applying them directly on the production systems. Netflix for example created a tool called Chaos Monkey~\cite{Izrailevsky2011TheArmyLocal} to randomly disable production servers and to observe how their platform would handle such failures. Chaos Monkey has inspired a number of similar approaches for different production environments, such as Kubernetes~\cite{kube-monkeyLocal,pod-reaperLocal,powerfulsealLocal}, Docker~\cite{AlexeiLedenev2017,dockerchaosmonkeyLocal}, AWS~\cite{chaos-lamdbaLocal} or private clouds~\cite{gomjabbarLocal} or even the JVM~\cite{Zhang2018AJVM} by forcing exceptions. 

\chaosorca is a novel Chaos Engineering approach, which proposes to perturb systems calls in a controlled manner. The blast radius of experiments conducted with \chaosorca is controlled by cgroups and namespaces, which makes it possible to use it in production: only targeted containers will be affected.

\subsection{Observability}

Monitoring is a well researched area. For example, PyMon is a lightweight monitoring solution to collect monitoring metrics from the open-source tool monit~\cite{GrobmannMarcel2017MCSa}; Dargos is a decentralized architecture for monitoring cloud environments with a low overhead~\cite{POVEDANOMOLINA20132041}

Moving to container-specific approaches, cAdvisor~\cite{cAdvisorLocal} by Google enables the collection and export of container metrics. It provides metrics about CPU, memory, disk and network usage, which can be for example used to decide when to scale in and out~\cite{CasalicchioEmiliano2017MDPW,CasalicchioEmiliano2017AoCT}. Node\textunderscore exporter~\cite{nodeexporterLocal} is a tool able to export many different metrics about hardware and operating system, examples include, CPU, disk, network and memory statistics. 

In Chaos Engineering, metrics are necessary to evaluate the steady state of the system (before pertubation) and the current state of the system during an experiment.
Netflix and Google have created a tool called Kayenta~\cite{kayentaLocal}, which does automated canary analysis by comparing the key metrics of the new deployment version to the old one. If the canary degrades metrics too much it can automatically abort the canary~\cite{kayentaAutomatedNetflixLocal}.

\chaosorca does not advocate for a specific monitoring approach provided that it can push its data into a Prometheus timeseries database. If not, an adapter needs to be implemented. Out of the box, \chaosorca integrates with cAdvisor to resources (CPU, RAM, etc), Bpftrace to monitor the invocation of system calls and PyShark to monitor HTTP requests.

\section{Conclusion}\label{sec:conclusion}
In this paper, we have presented \chaosorca, a novel Chaos Engineering approach to actively inject system call failures into a containerized application in order to evaluate its resilience in face of such perturbations. Under the hood, \chaosorca leverages Linux cgroups and namespaces to precisely control the blast radius of chaos experiments, \chaosorca also integrates with different and complementary monitoring solutions to increase the observability of the system. \chaosorca introduces perturbations by introducing errors or delays in the execution of the system calls, as specified in a resilience experiment.

Our evaluation on three different real-world applications shows that system call level perturbation is promising to detect weaknesses in self-protection and resilience mechanisms embedded in those applications. Our evaluation also shows that \chaosorca improves the observability of a containerized application without modifying any source code or configuration in the target container. 
In the future, we plan to make \chaosorca automatic in the context of a container orchestration framework such as Docker Swarm or Kubernetes.

\section*{Acknowledgements}
This work was partially supported by the Wallenberg AI, Autonomous Systems and Software Program (WASP) funded by the Knut and Alice Wallenberg Foundation.

\bibliographystyle{cas-model2-names}

\bibliography{references,references-local}

\end{document}